\documentclass[12pt,letterpaper]{article}
\pdfoutput=1
\usepackage{jheppub2}
\usepackage{amsmath}
\usepackage{bm}
\usepackage{comment}
\usepackage{rotfloat}
\usepackage{amssymb}

\hypersetup{
    colorlinks=true,       
    linkcolor=red,          
    citecolor=blue,        
    filecolor=magenta,      
    urlcolor=blue           
}

\def\tr{\,{\rm tr}\,}

\def\S{\mathcal{S}}
\def\C{\mathcal {C}}
\def\R{\mathcal{R}}
\def\P{\mathcal{P}}

\def\T{\mathcal{T}}
\def\tftp{p}

\newcommand{\ket}[1]{|#1 \rangle}

\preprint{INT-PUB-18-025}

\title{Dihedral symmetry in $SU(N)$ Yang-Mills theory}

\author[a]{Kyle Aitken,}
\author[b]{Aleksey Cherman,}
\author[c]{Mithat \"Unsal}
\emailAdd{kaitken17@gmail.com}
\emailAdd{aleksey.cherman.physics@gmail.com}
\emailAdd{unsal.mithat@gmail.com}
\affiliation[a]{Department of Physics, University of Washington, Seattle, WA 98105 USA}
\affiliation[b]{Institute for Nuclear Theory, University of Washington, Seattle, WA 98105 USA}
\affiliation[c]{Department of Physics, North Carolina State University, Raleigh, NC, USA}

\abstract{ 
We point out that charge conjugation and coordinate reflection symmetries do not commute with the center symmetry of $SU(N)$ YM theory when $N>2$.  As a result, for generic values of the $\theta$ angle, the group of discrete zero-form symmetries of YM theory on e.g. the spacetime manifold $\mathbb{R}^3\times S^1$ includes the  dihedral group $D_{2N}$ which is non-Abelian for $N>2$.  At $\theta = \pi$, the non-Abelian factor in the symmetry group is enhanced to $D_{4N}$ due to discrete 't Hooft anomaly considerations.  We illustrate these results in YM theory as well as in a simple quantum mechanical model, where 
we study representation theory as a function of $\theta$ angle. 
}

\begin{document}

\maketitle

\section{Introduction}
Internal symmetries are a familiar feature in quantum field theory with many established properties.  For example, places where symmetry realizations change can be associated with the emergence of gapless excitations.     Often, the realizations of internal symmetries are constrained by 't Hooft anomaly matching.  Additionally, in relativistic QFTs,  the Coleman-Mandula theorem \cite{Coleman:1967ad} implies that continuous internal symmetries commute with the Poincare group so that the full symmetry group of the theory, $G$, is a direct product: $G = G_{\rm Poincare} \times G_{\rm internal}$.

All of these features are illustrated in QCD.  QCD with $N\geq 3$ colors has a $G_{\rm internal} = [SU(N_F)_V \times SU(N_F)_A \times U(1)_Q]/\left( \mathbb{Z}_{N_F} \times \mathbb{Z}_{N}  \right)$ flavor symmetry in the chiral limit where $m_q = 0$, and $G_{\rm QCD} = G_{\rm Poincare} \times G_{\rm internal}$.   The $SU(N_F)_A$ part of the internal symmetry group has an 't Hooft anomaly.  This can be used to argue that when $m_q=0$ the low-energy effective theory describing fluctuations about the thermodynamic ground state must include some gapless degrees of freedom. For some values of $N_{F}$ and $N$, these gapless degrees of freedom are associated with spontaneous breaking of the $SU(N_F)_A$ symmetry.    

Here our focus will be on pure $SU(N)$ Yang-Mills theory
\begin{align}
S = \frac{1}{4g^2} \int d^4x~F^a_{\mu\nu}F^{a\mu\nu} + i \frac{\theta}{16\pi^2} \int d^4x~\epsilon^{\mu\nu\rho\sigma} F^a_{\mu\nu}F^a_{\rho\sigma} \, ,
\end{align} 
with $\mu,\nu = 1,\ldots,4$ and $a=1,\ldots,N$. Pure YM theory has no conventional internal symmetries which would act on local operators.  It has long been known, however, that it does have a subtler type of internal symmetry, $G_{\rm internal} = \mathbb{Z}_{N}$  center symmetry \cite{Polyakov:1978vu,Susskind:1979up,Gross:1980br,Weiss:1981ev}. 
Center symmetry acts non-trivially on certain line operators, but it does not act on  local operators.   
In the language of \cite{Gaiotto:2014kfa} center symmetry is a ``1-form symmetry", which can be contrasted with e.g. the chiral symmetry of QCD, which is a ``0-form symmetry" whose natural charged objects are local operators.  
It turns out that  just as with more familiar 0-form symmetries,  center symmetry can participate in 't Hooft anomalies \cite{Gaiotto:2017yup}.  
In particular there is a mixed 't Hooft anomaly between center symmetry and $CP$ symmetry at $\theta = \pi$ for even $N$, and a closely related notion of ``global inconsistency" for odd N \cite{Gaiotto:2017yup,Tanizaki:2017bam}.     
 
If the conclusions of the Coleman-Mandula theorem were to apply to center symmetry, then center symmetry would commute with $G_{\rm Poincare}$.  However, one cannot appeal to this theorem for two reasons.  First, the Coleman-Mandula theorem is derived for continuous internal symmetries, while the center symmetry of $SU(N)$ YM theory is discrete.  Second, the Coleman-Mandula theorem follows  from working out the constraints of symmetries on the  S-matrix for relativistic particle scattering, while the charged objects for center symmetry are associated to string-like extended operators.  
Indeed, we find that for pure $SU(N)$ YM theory on $\mathbb{R}^{3,1}$, the full symmetry group $G_{\rm YM}$ is generally \emph{not} a direct product:
\begin{align}
G_{\rm YM} \neq G_{\rm Poincare} \times G_{\rm internal}.
\end{align}
In particular, when $N\geq 3$ center symmetry transformations do not commute with a simultaneous transformation of parity and time reversal, $PT$, or with charge conjugation $C$.\footnote{This was noted but not explored in \cite{Aitken:2017ayq}.}  $PT$, $C$, and center transformations are symmetries of YM theory for all values of $g$ and $\theta$, so these two symmetries generate a discrete non-Abelian subgroup of $G^{\rm disc}_{\rm YM} \subset G_{\rm YM}$ for $N\geq 3$.  However, we will see that the nature of $G^{\rm disc}_{\rm YM}$ depends on $\theta$. 

We will show that when $SU(N)$ YM theory is compactified on $\mathbb{R}^3\times S^1$, the discrete 0-form symmetries fit into the group
\begin{align}
G^{\rm discrete}_{\rm YM} = \begin{cases}
D_{2N} \times \mathbb{Z}_2 \times \mathbb{Z}_2
& \theta=0\; \mathrm{mod}\; 2\pi \\
D_{4N} \times \mathbb{Z}_2 \times \mathbb{Z}_2
& \theta=\pi\; \mathrm{mod}\; 2\pi \\
D_{2N} & \text{otherwise}
\end{cases}
\label{eq:ym_theta_sym}
\end{align}
Here, $D_{2N}$ is the dihedral group of symmetries of a regular planar $N$-gon.  The dihedral group involves the 0-form part of center  symmetry, which acts on Wilson loops which wind around $S^1$, as well as charge conjugation.  The $\mathbb{Z}_2 \times \mathbb{Z}_2$ factors are related to parity and time-reversal symmetries.   Compactification on a circle simplifies the discussion but is not essential, see Sec. \ref{sec:symmetry} for a discussion concerning the symmetries on $\mathbb{R}^4$.

    \begin{figure}[t]
\begin{center}
\includegraphics[width = \textwidth]{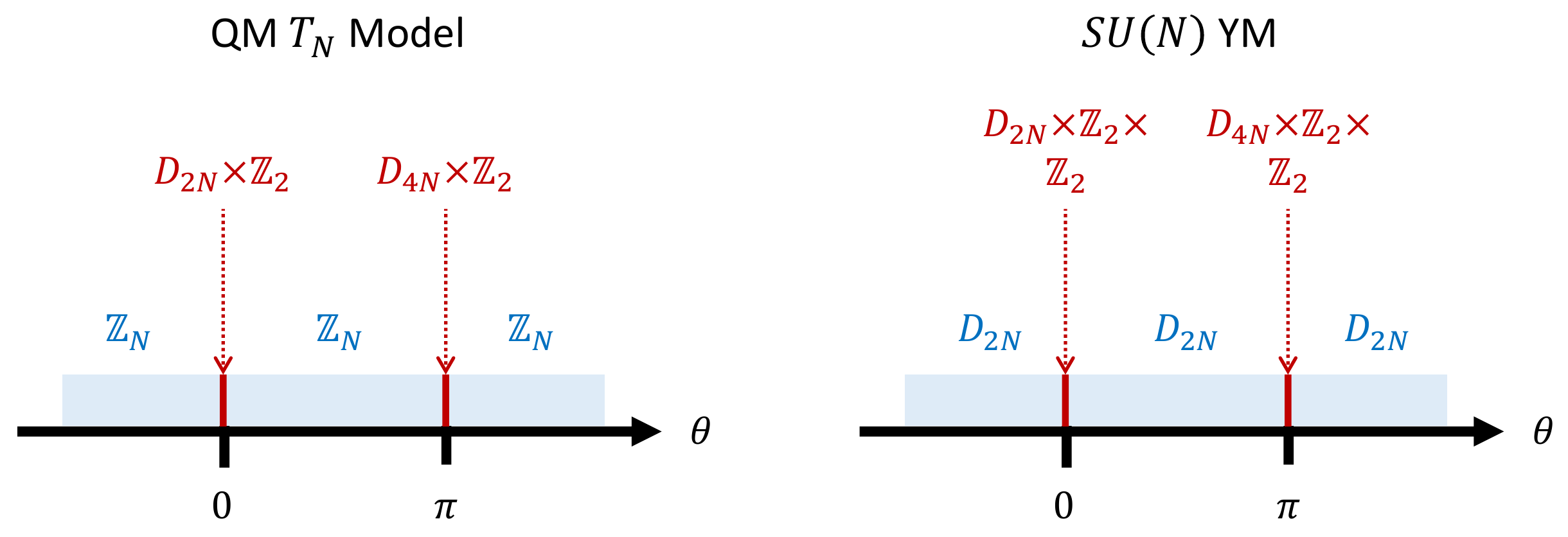}
\caption{[Color Online.]  A summary of the symmetries of $SU(N)$ YM theory (right) and of a related $T_N$ toy model from quantum mechanics (left), as a function of $\theta$.  }
\label{fig:symm_fig}
\end{center}
\end{figure}

The rest of the paper is concerned with illustrating how these symmetries behave in two different calculable settings.  First, we discuss a simple quantum-mechanical toy model in Sec.~\ref{sec:T_N_model} where many of the ideas can be appreciated in the simplest possible context.  In Sec.~\ref{sec:YM}, we then explore the symmetries of a calculable deformation of YM theory obtained by a compactifion on a small circle with stabilized center symmetry.  This semiclassically calculable regime was uncovered in \cite{Unsal:2008ch}, and intensively explored in related works, see e.g. 
\cite{
Unsal:2007vu,Unsal:2007jx,Shifman:2008ja,Shifman:2008cx,Shifman:2009tp,
Cossu:2009sq,Myers:2009df,Simic:2010sv,Unsal:2010qh,Azeyanagi:2010ne,Vairinhos:2011gv,
Thomas:2011ee,Anber:2011gn,Unsal:2012zj,
    Poppitz:2012sw,
    Poppitz:2012nz,Argyres:2012ka,Argyres:2012vv,
    Anber:2013doa,Cossu:2013ora,
    Anber:2014lba,Bergner:2014dua,Bhoonah:2014gpa,Li:2014lza,
    Anber:2015kea,Anber:2015wha,Misumi:2014raa,
    Cherman:2016hcd,Aitken:2017ayq,Anber:2017rch,Anber:2017pak,Anber:2017tug,Anber:2017ezt}.   A comparison of the symmetries between our QM toy model and $SU(N)$ YM theory is given in Fig.~\ref{fig:symm_fig}.  Our results are summarized in Sec.~\ref{sec:outlook}, and end with some appendices with details on some of our calculations.
    
In a companion paper \cite{Aitken:2018mbb}, we further explore the vacuum properties of the deformed YM theory as a function of $\theta$.

\section{Non-Abelian global symmetry of YM theory}
\label{sec:symmetry}
In this section we argue that the discrete part of the symmetry group of YM theory $G^{\rm discrete}_{\rm YM}$ includes the dihedral group $D_{2N}$.   This involves showing that center symmetry does not commute with charge conjugation $C$. Equivalently, center symmetry does not commute with $PT$ symmetry;  our discussion below will only explicitly refer to $C$ for simplicity. 

Since center symmetry does not act  on any local operators, a non-trivial check of the symmetry group generated by center symmetry and charge conjugation will involve consideration of line operators.   For simplicity of exposition, we work in Euclidean space.    We will first discuss the symmetries on $\mathbb{R}^3 \times S^1$, and then comment on the generalization to $\mathbb{R}^4$.

First, take spacetime to be $\mathbb{R}^3 \times S^1$, with $S^1$ the $x_4$ direction.  The zero-form part of center symmetry acts non-trivially on `Polyakov loops' --- Wilson loops wrapping the circle $\tr \Omega= \tr P \exp(i \oint dx_4 A_4)$, which are local with respect to $\mathbb{R}^3$.    
The action of center symmetry is 
\begin{align}
\S \cdot \tr \Omega = \omega \tr \Omega.
\end{align}
where the exponent of $\omega = e^{2\pi i /N}$ is the charge of $\tr \Omega$,  and we have denoted the operator implementing center symmetry transformations by $\S$. Of course, $\S^N=1$. 

The theory is invariant under charge conjugation symmetry $\mathcal{C}$ at arbitrary $\theta$-angle  since topological term 
respects   $\mathcal{C}$.
 Charge conjugation maps $\C: \Omega \to \Omega^{\dag}=\Omega^{-1}$ so that $\mathcal{C}^2=1$. Let us now work out the group obeyed by $\S$ and $\mathcal{C}$. One can then verify that
\begin{align}
\C \cdot \S \cdot \C \cdot \tr \Omega = \S^{-1} \cdot \tr \Omega.
\end{align}
Thus $\C$ and $\S$ do not commute.  In fact, they obey the defining relations of the dihedral group of symmetries of a regular planar $N$-gon,
\begin{align}
D_{2N}=\{\S,\,\C\; | \; \S^N = 1, \,  \C^2 = 1, \, \C \, \S \, \C = \S^{-1} \}.
\label{eq:d2N_algebra}
\end{align}

At $\theta = 0$ YM theory has parity $\mathcal{P}: x_{j} \to -x_{j}, j=1,2,3$ and $x_4$-reflection $\mathcal{R}: x_4 \to L-x_4$ symmetries.  There is also an $SO(3)$ Lorentz symmetry associated with the non-compact directions. It is easy to see that center symmetry also does not commute with $\R$ because its behavior when acting on $\tr \Omega$ is analogous to charge conjugation, $\R \cdot \tr \Omega = \tr \Omega^{-1}$.  But $[\S,\P] =[\C,\R]=[\C,\P] = 0$.  However, while $\R$ and $\P$ are manifestly symmetries at $\theta = 0$,  they are not symmetries for generic $\theta \neq 0, \pi$.  

At $\theta = \pi$ there is either a mixed 't Hooft anomaly or a global inconsistency between center and $CP$ symmetries \cite{Gaiotto:2017yup}, depending on whether $N$ is even or odd. Assuming that center symmetry is not spontaneously broken for all $\theta$, when there is mixed 't Hooft anomaly at $\theta = \pi$ there are two possibilities for the vacuum structure: (1) $CP$ is spontaneously broken or (2) there is a nontrivial topological field theory which matches the anomaly in the IR limit. A global inconsistency condition at $\theta = \pi$ is slightly weaker, and in addition to the two options above, can also be satisfied if there are phase transitions away from $\theta = 0,\pi$ \cite{Gaiotto:2017yup,Tanizaki:2017bam}. 

Especially for large $N$, spontaneous breaking of $CP$ at $\theta = \pi$ seems like the most probable way these anomaly/inconsistency conditions would be satisfied, and we assume this is the case in writing expressions in the $\mathbb{R}^4$ limit.  On $\mathbb{R}^3 \times S^1$, $CP$ breaking can be shown explicitly. We demonstrate the  anomaly/global inconsistency conditions at $\theta  = \pi$ imprint themselves on the symmetry group by leading to a central extension.  So at  $\theta = \pi$, the discrete global symmetry contains a factor of $D_{4N}$, the double-cover of $D_{2N}$.\footnote{The fact that the symmetry group of $SU(2)$ YM theory involves a $D_8$ factor at $\theta = \pi$ was discussed in \cite{Gaiotto:2017yup}.}  Taken together, these considerations imply the claim from the introduction in \eqref{eq:ym_theta_sym}.  

Note that the dihedral group $D_{4}$ is isomorphic to the Abelian group $\mathbb{Z}_2 \times \mathbb{Z}_2$, but $D_{2N}$ is non-Abelian for all $N>2$.  So when $N=2$ the discrete symmetry group of YM theory is Abelian for $\theta \neq \pi$, and becomes non-Abelian only when $\theta =\pi$.   However, for all $N > 2$,  $G^{\rm discrete}_{\rm YM}$ is non-Abelian for all $\theta$, and the group of zero-form symmetries $\S, \P, \R, \C$ will be shown to take the form \eqref{eq:ym_theta_sym}.

We now turn our attention to $\mathbb{R}^4$. Here it is helpful to adopt the language of \cite{Gaiotto:2014kfa}, in which center symmetry is viewed as a $p$-form symmetry with $p=1$.  The charges of $p$-form symmetries are measured by integrating conserved $d-p-1$ currents on closed $d-p-1$-dimensional manifolds,  and are associated to charges of operators supported on manifolds of dimension $p$.  The charge of such an operator is non-zero when its worldvolume manifold has non-vanishing linking number with some $d-p-1$-dimensional manifold where one puts the operator generating the symmetry.   

In the case of $1$-form center symmetry, the basic charged operators are Wilson lines with appropriate topological properties. In particular, consider an open Wilson line defined on a curve $\gamma$ whose ends go off to infinity in different directions, for instance along $x_4 \to \pm \infty$.  One can think of such a line operator $\Omega(\gamma)$ as being associated with a  probe fundamental quark-anti-quark pair with separation taken to infinity, with e.g. the quark going to $x_4 \to +\infty$ and the anti-quark going to $x_4 \to -\infty$. (If the $x_4$ direction is compactified to $S^1$, this open Wilson line becomes precisely the Polyakov loop considered earlier.)   For our purposes, it will be useful to associate the operator generating center symmetry with the closed two-dimension surface $\Sigma_2$ which spans the $x_1$-$x_2$ plane. In this case, center symmetry acts on $\tr \Omega(\gamma)$ as \cite{Gaiotto:2014kfa}
\begin{align}
\S \cdot \tr\Omega(\gamma) = \omega^{\ell (\gamma, \Sigma_2) } \tr\Omega(\gamma) =  \omega^{+1} \tr\Omega(\gamma) \,.
\end{align}
where $ \ell (\gamma, \Sigma_2) $ is the linking number of $\gamma$ with $\Sigma_2$  \cite{Gaiotto:2014kfa},  which is  
$+1$ in the case above. 

Now consider charge conjugation. This symmetry interchanges quarks and anti-quarks, so it acts on $\Omega(\gamma)$ as
\begin{align}
\mathcal{C} \cdot \tr \Omega(\gamma)= \tr\Omega(\gamma)^{\dag}= \tr \Omega(\gamma^{-1})
\end{align}
so $\mathcal{C}$ flips the orientation of $\gamma$. Flipping the orientation of $\gamma$ flips the sign of the linking number of $\gamma$ with $\Sigma_2$, $ \ell (-\gamma, \Sigma_2) = - \ell (\gamma, \Sigma_2) $.  The operator group then follows as before,
\begin{align}
\C \cdot \S \cdot \C \cdot  \tr\Omega(\gamma) = \omega^{-1}  \tr\Omega(\gamma)  = \S^{-1} \cdot  \tr\Omega(\gamma)  .
\end{align}
Thus $\C$ and $\S$ do not commute on $\mathbb{R}^4$.  It is also easy to see that $\S$ does not commute with $\R$, the $\theta = 0, \pi$ symmetry operator which now maps $x_4 \to -x_4$.  

Rather trivially, symmetries of quantum systems can be associated with groups because, given some state $\ket{\psi}$ in Hilbert space which transforms non-trivially under a symmetry, one can verify that the symmetry action obeys the group axioms.  In our case, choosing $\ket{\psi}=\tr\Omega \ket{0}$ our remarks above imply that the actions of the $\C$ and $\S$ transformations obey the group axioms, and combine into the symmetry group $D_{2N}$.  Nevertheless, we are dealing with the somewhat unusual situation of considering the combination of a 0-form symmetry and a 1-form symmetry.  As this paper was being prepared for submission, Refs.~\cite{Cordova:2018cvg,Benini:2018reh} appeared (see also e.g. Ref.~\cite{Kapustin:2013uxa}), where it is argued that the general algebraic structure appropriate to discuss the mixture of $0$-form and $1$-form symmetries is a ``$2$-group"\cite{BaezLauda}.~\footnote{We are grateful to K.~Jensen and S.~Gukov for discussions on this point.}  

\subsection{Physical consequences}
We now comment on some physical consequences of the existence of the dihedral non-Abelian symmetry in $SU(N)$ YM theory.  The fact that charge conjugation and center symmetry do not commute means that the associated charge operators cannot be simultaneously diagonalized.  This means that if one considers a state that transforms non-trivially under both center and charge conjugation symmetry, one cannot simultaneously specify its center symmetry and charge conjugation quantum numbers.  Of course, this means that the existence of the $D_{2N}$ symmetry does \emph{not} imprint itself on the correlation functions of local operators.  One must consider correlation functions of appropriate line operators to see the symmetry.  

For example, consider $SU(N)$ YM theory with $N>2$ on $\mathbb{R}^3\times S^1$.  Finite-energy states transforming under center symmetry can be built out of Wilson loops wrapping $S^1$.  Then one can consider scattering amplitudes involving such states, for example at $\theta = 0$.  Suppose we choose to specify the center labels of the states.    Then the fact that one cannot simultaneously specify the center and charge conjugation quantum numbers --- which is due to the existence of the $D_{2N}$ symmetry --- means that one has to sum over the $C$ quantum numbers for both incoming and outgoing states in computing the scattering amplitudes.  

At high temperature, center symmetry is spontaneously broken in pure YM theory.  It would be interesting to understand the physical implications of the non-commutativity of center symmetry and e.g. $PT$ symmetry in this setting.

\section{Dihedral symmetries in a quantum mechanical model}
\label{sec:T_N_model}

As a warm up for studying the symmetries and dynamics of $SU(N)$ gauge theory as a function of $\theta$, we will  first consider the quantum mechanical system of a particle on a circle,  $q(t) = q(t)+2\pi$, in the presence of a potential with $N$ degenerate minima. This class of models  are referred to as  $T_{N}$ models in  \cite{Unsal:2012zj}, where their non-perturbative properties were examined semi-classically. 
  The Euclidean action of the model is
\begin{align}
S_{T_N}(g,\theta) = \frac{1}{g^{2}}\int dt\,\left[\frac{1}{2}\dot{q}^{2}-\cos\left(Nq\right)\right]-i\frac{\theta}{2\pi}\int dt\,\dot{q}.
\label{eq:tN_definition}
\end{align}
The potential has $N$ degenerate minima at $q_n = \tfrac{2\pi n}{N}, n = 0,1, \cdots N-1$.  But the system does not have $N$ degenerate ground states: tunneling/instanton effects typically lift the degeneracies seen in perturbation theory. However, this does not mean that the ground state is always unique.  For some values of $\theta$, it turns out to be doubly degenerate. We discuss the ground state structure below from a perspective that we will find useful in YM theory.

Analogies between the 1d $T_N$ model and 4d $SU(N)$ YM theory were previously explored in \cite{Unsal:2012zj}, and a detailed analysis of the symmetries of a very closely related model appeared in \cite{Kikuchi:2017pcp}.  The discussion in  Sec.~\ref{sec:QM_algebra} thus has  overlap with \cite{Kikuchi:2017pcp}, but the subsequent representation-theoretic perspective presented in Sec.~\ref{sec:representation_theory_QM} is new.   A discussion of the symmetries of the $T_2$ model as a function of $\theta$ appeared in an appendix of \cite{Gaiotto:2017tne}, but our focus will be on features that appear once $N>2$.   Also, a discussion of 't Hooft anomalies from the path integral perspective is given in Appendix~\ref{appendix:path_integral_QM}.  This material in this Appendix closely follows the presentation of  \cite{Kikuchi:2017pcp}, and we include it here for completeness.

\subsection{Symmetry group as a function of $\theta$}
\label{sec:QM_algebra}

Consider the symmetry group of the $T_N$ theory.  Classically, there is a shift symmetry $\mathcal{S}$ as well as `charge conjugation' $\mathcal{C}$ and `time reversal' $\mathcal{T}$ symmetries acting as
\begin{align}
&\mathcal{S}: \; q \to q - 2\pi/N \\
&\mathcal{C}: \; q \to -q\\
&\mathcal{T}: \; t \to - t.
\end{align}
In the quantum theory, the shift symmetry can represented by the operator
\begin{align}
\S = e^{\tfrac{2\pi i}{N} \hat{p}} \, .
\end{align}
where $\hat{p}$ is the momentum operator obeying the canonical commutation relation $[\hat{q}, \hat{p}] = i$.  As befits a symmetry operator,  $\S$ commutes with the Hamiltonian
\begin{align}
\hat{H} = \frac{1}{2} \left(\hat{p} - \tfrac{\theta}{2\pi}\right)^2 - \cos(N \hat{q}) \, .
\end{align}    
Demanding that $\T$ and $\C$ leave the Hamiltonian invariant in e.g. the coordinate basis, one sees that for $\theta = 0$, the $\T$ and $\C$ symmetries both act by sending $\hat{p} \to -\hat{p}$, while at $\theta = \pi$, the $\T$ and $\C$ symmetries both act by sending $\hat{p} \to - \hat{p} +1$. Thus  e.g. $\T \hat{p} \T^{-1} = -\hat{p}$ at $\theta = 0$, but  $\T \hat{p} \T^{-1} = -\hat{p}+1$ at $\theta = \pi$. One can check that at both $\theta = 0$ and $\theta = \pi$, the $\C$ and $\T$ operators commute, 
\begin{align}
[\T,\C] = 0.
\end{align}

In Minkowski space time-reversal is an anti-unitary operation, so in addition to sending $t \to -t$, $\T$ acts by complex conjugation $\T i \T^{-1}=-i$, in contrast to $\C$, which is unitary. One can check that this implies that $\left[ \T, \S \right]=0$ in Minkowski space.

$\C$ does not commute with $\S$ at $\theta = 0$. To see this, note the lowest-lying states of the system can be thought of being associated with nodeless wavefunctions $\ket{q_n}$ localized near the $N$ minima.  These states are called Wannier states. From $\ket{q_n}$ one can build states with good quantum numbers $\ket{k}$ under $\S$ by a discrete Fourier transform 
\begin{align}
\ket{k} = \frac{1}{\sqrt{N}} \sum_{n = 0}^{N-1} \omega^{-n k} \ket{q_n}.
\label{eq:bloch_states}
\end{align}
The states $\ket{k}$ are called  Bloch states.  Then
\begin{align}\label{eq:tn_S_trans}
\S \ket{k} = \omega^{k} \ket{k}
\end{align}
with $\ket{k}=\ket{k\text{ mod }N}$. 
Then one can check that
\begin{align}\label{eq:tn_C_trans}
\mathcal{C} \ket{k} = \begin{cases}
\ket{N-k}
& \theta=0\\
\ket{N-k+1}
& \theta=\pi
\end{cases}.
\end{align}
As a result, at $\theta = 0$ the symmetry operators obey the group
\begin{align}
\C \S \C^{-1} =
\S^{-1} \,. 
\label{eq:tn_zero_algebra}
\end{align}
Given that $\T^2 = \C^2 = 1$, the complete symmetry group is isomorphic to
\begin{align}
G^{\theta = 0}_{T_N} = D_{2N} \times \mathbb{Z}_2.
\end{align}

On the other hand,  at  $\theta =\pi$ we instead obtain
\begin{align}
\C \S \C^{-1} =
\omega^{-1}\S^{-1} \,. 
\label{eq:tn_pi_algebra}
\end{align}
The appearance of the factor $\omega^{-1}$ on the right-hand side means that the group is not closed in terms of $\C, \T$ and $\S$.  This is a symmetry-group-level indication of a 't Hooft anomaly or global inconsistency between these symmetries. As a result, one  of these symmetries  must be spontaneously broken at $\theta = \pi$, or there must be a phase transition at some  $\theta$ between $0$ and $\pi$.%
\footnote{To decide which of these two symmetries are `actually' broken, it is helpful to note that there is no way to explicitly break $\T$ at $\theta = k = 0$ while preserving $\S$.  But if we change the potential 
\begin{align}
V = \cos[N q] \to \cos[N (q + \alpha)],
\end{align}
then for any fixed $\alpha \neq 0$, the $\C$ symmetry $q \to -q$ is explicitly broken, but $\S$ and $\T$ are preserved.  One can then verify that this $\T$ and $\S$ remain globally inconsistent at $\theta = \pi$, so that one of them must be broken, and this turns out to be $\T$ \cite{Kikuchi:2017pcp}.  Then taking $\alpha$ to 0, we conclude that it is the $\T$ symmetry which is spontaneously broken at $\theta = \pi$ in our variant of the $T_N$ model defined by \eqref{eq:tN_definition}.}

One can try to redefine the operators to get a closed group, for example by 
$\tilde{\S}\equiv\omega^{\tftp}\S$.  We will refer to $p$ as a Chern-Simons coefficient, since this is how it appears in a path integral description of this system, see \cite{Gaiotto:2017yup,Kikuchi:2017pcp} and Appendix \ref{appendix:path_integral_QM}.  This will not spoil the relation $\tilde{\S}^{N}=1$ so long as $\tftp\in\mathbb{\mathbb{Z}}\,\,\text{mod}\,N$. With such a 
redefinition \eqref{eq:tn_pi_algebra} becomes
\begin{align}
\C \tilde{\S} \C^{-1} =
\omega^{2\tftp-1}\tilde{\S}^{-1} \,. 
\label{eq:tn_redefined_algebra}
\end{align}
To keep \eqref{eq:tn_redefined_algebra} isomorphic to \eqref{eq:tn_zero_algebra} requires $2\tftp-1=0\,\text{mod}\,N$.  

Now consider the case of odd and even $N$ separately. For even $N$, there is no solution to $2\tftp-1=0\,\,\text{mod\,}N$ for
$\tftp\in\mathbb{\mathbb{Z}}$.  Nevertheless, we can get a closed group by taking $\tftp=1/2$.  In the path integral description, this gives a Chern-Simons term with an improperly-quantized coefficient.  This is associated with a mixed 't Hooft anomaly. In the operator description, the choice $\tftp = 1/2$ gives 
\begin{equation}
\mathcal{C}\tilde{\S}\mathcal{C}^{-1}=\tilde{\S}^{-1}.\label{eq:restored alg tnm}
\end{equation}
But now the operator $\tilde{\S}$ satisfies
\begin{equation}
\tilde{\S}^{N}=-1,\qquad\tilde{\S}^{2N}=1 .
\end{equation}
As a result, the symmetry group is now isomorphic to $D_{4N} \times \mathbb{Z}_2$, the central extension of $D_{2N}\times \mathbb{Z}_2$.\footnote{One can think of $D_{4N}$ as the spin group of $D_{2N}$, in the sense that under a $2 \pi$ shift of $q$ (which is a rotation in target space), states goes to minus themselves, and only go back to themselves under a $4 \pi$ shift.}  The central extension is the operator-group realization of the anomaly. 

For odd $N=2m-1$, $2\tftp-1=0\,\,\text{mod}\,\,N$ is satisfied with the choice $\tftp=\left(N+1\right)/2$. Hence, if we define $\tilde{\S}\equiv\omega^{\left(N+1\right)/2}\S$,
this also reduces to (\ref{eq:restored alg tnm}) since
\begin{align}
\tilde{\S}^{N} & =\omega^{\left(N+1\right)N/2}\S^{N}=\omega^{2mN/2}\left(\omega^{N}\right)^{m}=1.
\end{align}
However, if we insist on preserving the $D_{2N}\times\mathbb{Z}_2$ symmetry at $\theta = 0$, then we must choose the value of Chern-Simons coefficient to be $\tftp=0$ (i.e. the original operator definition). This is the manifestation of the inconsistency condition and results in a centrally extended group $D_{4N} \times \mathbb{Z}_2$ at $\theta = \pi$.  

Collecting our results, the symmetry group of the $T_N$ model as a function of $\theta$ is isomorphic to 
\begin{equation}
G_{T_N} = \begin{cases}
D_{2N} \times \mathbb{Z}_2 
& \theta=0\\
D_{4N} \times \mathbb{Z}_2
& \theta=\pi\\
\mathbb{\mathbb{Z}}_{N} & \text{otherwise}.
\end{cases}
\end{equation}

\subsection{Representations of the dihedral group for $\protect\theta=0$ and $\protect\theta=\pi$}
\label{sec:representation_theory_QM}

We now explain how the states of the $T_N$ model fit into the representations of the dihedral group. The value of this discussion is that it relies on the symmetry group structure, rather than the underlying physics, and thus can later be applied almost verbatim to the YM case.

One can construct the $N$-dimensional representation of a dihedral group based on the behavior of the $N$ vacua of the $T_N$ model under the action of charge conjugation (equivalently, time-reversal) and  ${\mathbb Z}_N$ shift symmetry. The decomposition of this representation into irreducible representations (irreps) will show us the form of the energy spectrum and provide us another means to see how the degeneracy of the ground state changes between $\theta=0$ and $\pi$. For both $D_{2N}$ and $D_{4N}$, we find results consistent with the operator analysis above.
\begin{table}
\begin{centering}
\begin{tabular}{|c|c|c|c|c|c|c|c|c|}
\hline 
 & $1\left\{ 1\right\} $ & $2\left\{ r^{\pm1}\right\} $ & $2\left\{ r^{\pm2}\right\} $ & $\cdots$ & $2\left\{ r^{\pm\left(k-1\right)}\right\} $ & $1\left\{r^{k}\right\}$ & $k\left\{ sr^{2b}\right\} $ & $k\left\{ sr^{2b-1}\right\} $\tabularnewline
\hline 
${\bf A_{1}}$ & $1$ & $1$ & $1$ & $\cdots$ & $1$ & $1$ & $1$ & $1$\tabularnewline
${\bf A_{2}}$ & $1$ & $1$ & $1$ & $\cdots$ & $1$ & $1$ & $-1$ & $-1$\tabularnewline
${\bf B_{1}}$ & $1$ & $-1$ & $1$ & $\cdots$ & $\left(-1\right)^{k-1}$ & $\left(-1\right)^{k}$ & $1$ & $-1$\tabularnewline
${\bf B_{2}}$ & $1$ & $-1$ & $1$ & $\cdots$ & $\left(-1\right)^{k-1}$ & $\left(-1\right)^{k}$ & $-1$ & $1$\tabularnewline
${\bf E_{1}}$ & $2$ & $2c_{1}$ & $2c_{2}$ & $\cdots$ & $2c_{k-1}$ & $2c_{k}$ & $0$ & $0$\tabularnewline
${\bf E_{2}}$ & $2$ & $2c_{2}$ & $2c_{4}$ & $\cdots$ & $2c_{2k-2}$ & $2c_{2k}$ & $0$ & $0$\tabularnewline
$\cdots$ & $\cdots$ & $\cdots$ & $\cdots$ & $\cdots$ & $\cdots$ & $\cdots$ & $\cdots$ & $\cdots$\tabularnewline
${\bf E_{k-1}}$ & $2$ & $2c_{k-1}$ & $2c_{2k-2}$ & $\cdots$ & $2c_{\left(k-1\right)^{2}}$ & $2c_{\left(k-1\right)k}$ & $0$ & $0$\tabularnewline
\hline 
\end{tabular}
\par\end{centering}
\caption{Character table for $D_{2M}=D_{2(2k)}$. Here, $c_{n}=\cos\left(\frac{2\pi n}{M}\right)$. The first row shows the number of elements in the respective conjugacy classes.
\label{tab:D2N_char_tab}}
\end{table}

Let us start by briefly reviewing a few properties of dihedral groups. A more detailed review and discussion of our results is given in Appendix \ref{app:Reps_dihedral}. We will work with a standard presentation of the dihedral group $D_{2M}$, which is given by
\begin{equation}
D_{2M}=\{ r,s|r^{M}=s^{2}=1,srs^{-1}=r^{-1}\}.
\end{equation}
The representations of this group differ for even and odd $M$, so we will consider the two cases separately in what follows. 

\begin{figure}[t]
\begin{center}
\includegraphics[width = \textwidth]{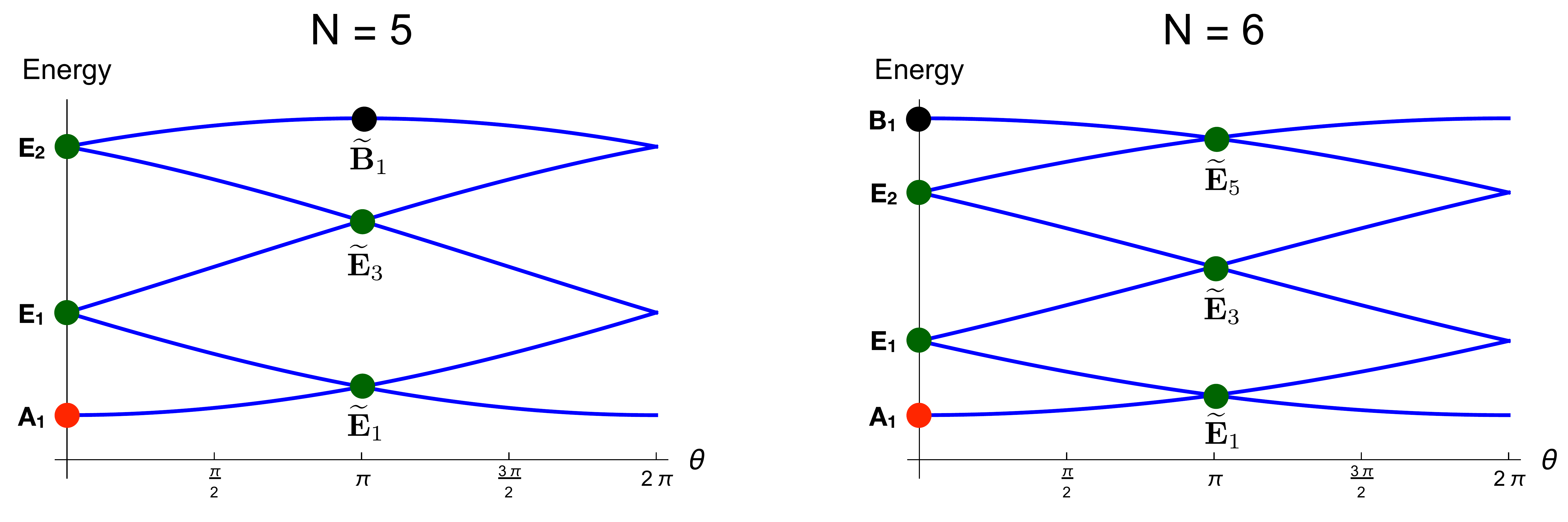}
\caption{An illustration of the energy levels of the $T_N$ model for $N=5$ and $N=6$.  At $\theta=0$, the ground state is unique, and fits into the one-dimensional ${\bf A_{1}}$ representation of $D_{2N}$, while the excited states fit into either the ${\bf E_{k}}$ representations (which are all two-dimensional) or into the ${\bf B_{1}}$ representation, which is one-dimensional.  At $\theta = \pi$, on the other hand, the ground state is always in the two-dimensional $\bf {\widetilde{E}
}_{1}$ representation of $D_{4N}$.}
\label{fig:reps_QM_fig}
\end{center}
\end{figure}

\begin{figure}[h]
\begin{center}
\includegraphics[width = \textwidth]{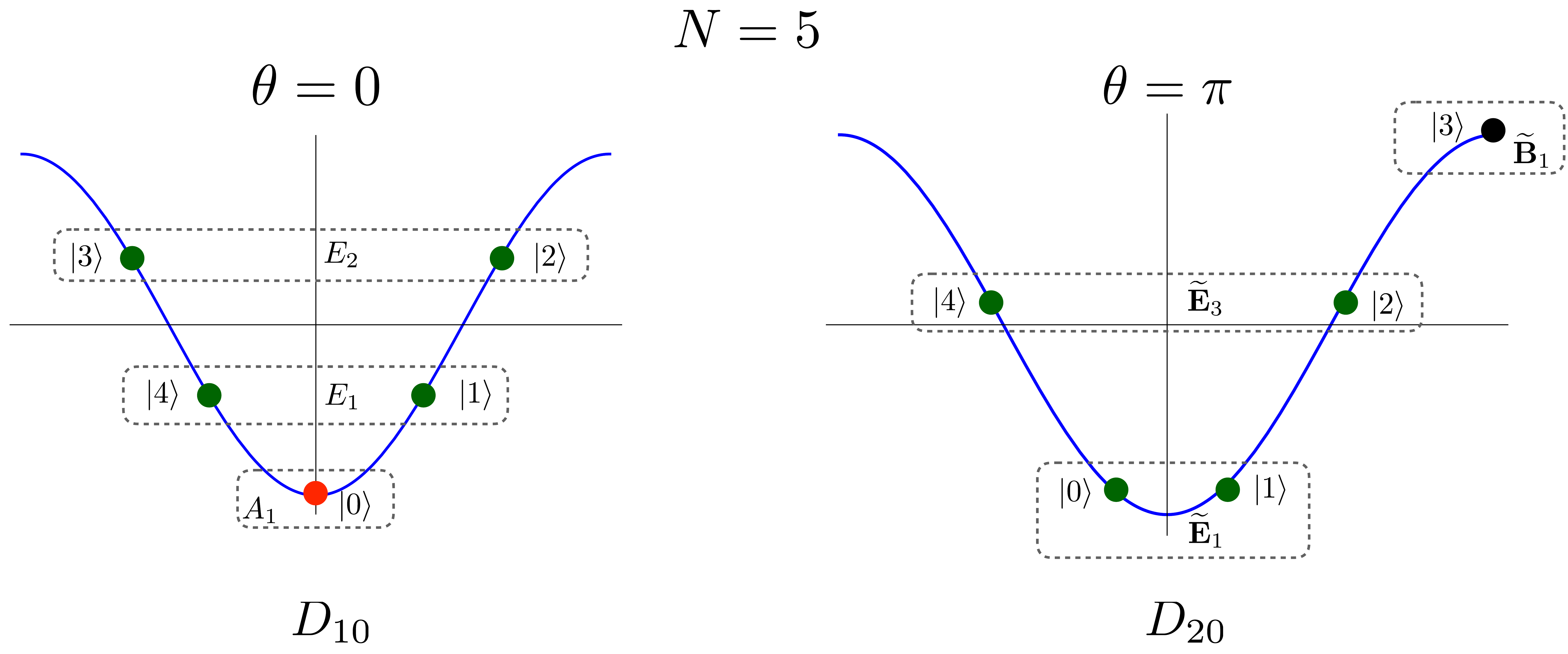}
\caption{A sketch of how the states of the $T_N$ model with $N=5$ and $\theta = 0$ and $\theta =\pi$ fit into the dihedral group $D_{10}$ and $D_{20}$ representations.  The Bloch states $\ket{k}$ are defined in \eqref{eq:bloch_states}. }
\label{fig:reps_QM_fig_2}
\end{center}
\end{figure}

Below, we will consider the representations which correspond to the low-lying states of the $T_{N}$ model, i.e. the $N$-low lying Bloch states $\ket{k}$, for the cases of $N$ even and odd.  Our goal is to understand the representation of the $N$-low lying states.  The results are visually summarized in Figs.~\ref{fig:reps_QM_fig} and \ref{fig:reps_QM_fig_2}, which plots the energies of these states as 
as function of $\theta$ angle, and are compatible with mixed anomalies/global inconsistencies as well as semi-classics.

\vspace{0.3cm}
\noindent
{\bf Even $N$:} 
For  $M=N=2k$, the $k+3$ conjugacy classes are\\
\begin{align}
\left\{ 1\right\} ,\left\{ r^{\pm1}\right\} ,\left\{ r^{\pm2}\right\} ,\ldots,\left\{ r^{\pm\left(k-1\right)}\right\} ,\left\{ r^{k}\right\} ,\left\{ sr^{2b}|b=1,\ldots,k\right\} ,\left\{ sr^{2b-1}|b=1,\ldots,k\right\} \label{eq:d_even cc tnm}
\end{align}
where the number of elements in the conjugacy classes are given by
\begin{equation}
\{1,\underset{k-1}{\underbrace{2,2,\ldots,2}},1,k,k\}.\label{eq:d_even cc size tnm}
\end{equation}
A character table for the representations of $D_{2M}$ is given in Table \ref{tab:D2N_char_tab}.

At $\theta=0$, the $N$ low lying states labeled by $\ket{k}$ transform under the action of $D_{2N}$ group elements. The conjugacy classes and number of elements in each class are given by (\ref{eq:d_even cc tnm}) and (\ref{eq:d_even cc size tnm}) with $M=N$.  It is straightforward to construct the $N$-dimensional representation associated with $N$-low lying states 
under the actions of $\S$ and $\C$. $\S$ simply introduces a vacuum-dependent phase to each of the states while $\C$ permutes them. The characters corresponding to the conjugacy classes listed in (\ref{eq:d_even cc size tnm}) are
\begin{equation}
\chi_{\text{even}}^{\theta=0}=\{N,\underset{k-1}{\underbrace{0,0,\ldots,0}},0,2,0\}.
\end{equation}
Character orthogonality then gives the decomposition in terms of irreps\\
\begin{equation} \label{eq:d_even th=0}
R_{\text{even}}^{\theta=0}={\bf A}_{1}\oplus {\bf E_{1}}\oplus {\bf E_{2}}\oplus\ldots\oplus {\bf E_{k-1}}\oplus {\bf B_{1}}
\end{equation}
where ${\bf A_{1}}$ and ${\bf B_{1}}$ are one-dimensional irreps and ${\bf E_{i}}$ is a 2 dimensional irrep (a doublet). ${\bf A_{1}}$ represents the unique ground state of this system which transform trivially under all group operations. 

At $\theta=\pi$, per our results of the previous subsection, the symmetry group is now $D_{4N}$. However, we should still construct an $N$-dimensional representations which tells us how the $N$ vacua now transform under this centrally extended group. The characters of this representation are\\
\begin{equation}
\chi_{\text{even}}^{\theta=\pi}=\{N,\underset{N-1}{\underbrace{0,0,\ldots,0}},-N,0,0\}.
\end{equation}
The decomposition in terms of irreps is now given by
\begin{equation} \label{eq:d_even th=pi}
R_{\text{even}}^{\theta=\pi}={\bf \tilde{E}_{1}}\oplus {\bf \tilde{E}_{3}}\oplus\ldots\oplus {\bf \tilde{E}_{2k-1}},
\end{equation}
(with ${\bf \tilde{E}_i}$ now irreps of $D_{4N}$). The fact that the ground state exhibits two-fold degeneracy in this simple quantum mechanics example is a manifestation of the \textquoteright t Hooft anomaly between $\mathbb{\mathbb{Z}}_{N}$ and $\mathbb{\mathbb{Z}}_{2}$ and is tied with the spontaneous breaking of the $\mathbb{\mathbb{Z}}_{2}$ symmetry.

\vspace{0.3cm}
\noindent
{\bf Odd $N$:} 
For odd $M=2k+1$, the $k+2$ conjugacy classes are
\begin{equation}
\left\{ 1\right\} ,\left\{ r^{\pm1}\right\} ,\left\{ r^{\pm2}\right\} ,\ldots,\left\{ r^{\pm k}\right\} ,\left\{ sr^{b}|b=1,\ldots,M\right\} \label{eq:d_odd cc tnm}
\end{equation}
where the number of elements in each conjugacy class is now
\begin{equation}
\{1,\underset{k}{\underbrace{2,2,\ldots,2}},N\}.\label{eq:d_odd cc size tnm}
\end{equation}
The corresponding character table is given in Table \ref{tab:D2N+1 char tab tnm}.

\begin{table}
\begin{centering}
\begin{tabular}{|c|c|c|c|c|c|c|}
\hline 
 & $1\left\{ 1\right\} $ & $2\left\{ r^{\pm1}\right\} $ & $2\left\{ r^{\pm2}\right\} $ & $\cdots$ & $2\left\{ r^{\pm k}\right\} $ & $N\left\{ sr^{2b}\right\} $\tabularnewline
\hline 
${\bf A_{1}}$ & $1$ & $1$ & $1$ & $\cdots$ & $1$ & $1$\tabularnewline
${\bf A_{2}}$ & $1$ & $1$ & $1$ & $\cdots$ & $1$ & $-1$\tabularnewline
${\bf E_{1}}$ & $2$ & $2c_{1}$ & $2c_{2}$ & $\cdots$ & $2c_{k}$ & $0$\tabularnewline
${\bf E_{2}}$ & $2$ & $2c_{2}$ & $2c_{4}$ & $\cdots$ & $2c_{2k}$ & $0$\tabularnewline
$\cdots$ & $\cdots$ & $\cdots$ & $\cdots$ & $\cdots$ & $\cdots$ & $\cdots$\tabularnewline
${\bf E_{k}}$ & $2$ & $2c_{k}$ & $2c_{2k}$ & $\cdots$ & $2c_{k^{2}}$ & $0$\tabularnewline
\hline 
\end{tabular}
\par\end{centering}
\caption{Character table for $D_{2M}=D_{2\left(2k+1\right)}$. Here, $c_{m}=\cos\left(\frac{2\pi m}{M}\right)$.
\label{tab:D2N+1 char tab tnm}}
\end{table}

 At $\theta=0$, the $N$ low lying states transform under the action of $D_{2N}=D_{2\left(2k+1\right)}$ group elements. The characters of the $N$-dimensional representation are given by:\\
\begin{equation}
\chi_{\text{odd}}^{\theta=0}=\{N,\underset{k}{\underbrace{0,0,\ldots,0}},1\}.
\end{equation}
In this case, the decomposition is given by\\
\begin{equation} \label{eq:d_odd th=0}
R_{\text{odd}}^{\theta=0}={\bf A_{1}}\oplus {\bf E_{1}}\oplus {\bf E_{2}}\oplus\ldots\oplus {\bf E_{k}}
\end{equation}
where ${\bf A_{1}}$ is again a one-dimensional irrep and ${\bf E_{i}}$ are 2 dimensional irreps of $D_{2N}$ with $N$ odd. ${\bf A_{1}}$ again represents the unique ground state of this system at $\theta=0$.

For the $T_{N}$ model with odd $N$, there is a global inconsistency condition at $\theta=\pi$ between $\S$ and $\T$ (or $\C$).  As a result, the vacuum cannot remain trivial which indicated either a non-trivial vacuum or a phase transition between $\theta=0$ and $\pi$. We will assume that the inconsistency implies the former such that the $N$ low lying states transform under the action of $D_{4N}=D_{2\left(4k+2\right)}$. This is the group that will give rise to the two-fold degenerate ground state we find as a result of the global inconsistency condition. We also assume the central extension comes about in the same manner as the even $N$ case, where $\tilde{\S}\equiv\omega^{1/2} \S$. Semi-classical instanton analysis  \cite{Unsal:2012zj} and  numerical diagonalization of \eqref{eq:tN_definition} found in e.g. \cite{Kikuchi:2017pcp} support the resulting degeneracies and $\theta$-dependence from this assumption.

 At $\theta=\pi$, the $N$ low lying states transform under the action of $D_{4N}=D_{2\left(4k+2\right)}$ group elements.
The characters of the conjugacy classes in this case are
\begin{equation}
\chi_{\text{odd}}^{\theta=\pi}=\{N,\underset{N-1}{\underbrace{0,0,\ldots,0}},-N,1,-1\}
\end{equation}
and the corresponding decomposition is
\begin{equation} \label{eq:d_odd th=pi}
R_{\text{odd}}^{\theta=\pi}={\bf \tilde{E}_{1}}\oplus {\bf \tilde{E}_{3}}\oplus\ldots\oplus {\bf \tilde{E}_{2k-1}}\oplus {\bf \tilde{B}_{1}}.
\end{equation}
${\bf \tilde{E}_{1}}$ denotes the ground state and exhibits two-fold degeneracy. Other ${\bf \tilde{E}}$-states are excited states, and ${\bf \tilde{B}_{1}}$ is the highest energy state (of the low lying states), which is a singlet.


\section{Dihedral symmetries in Yang-Mills theory on $\mathbb{R}^3\times S^1$}
\label{sec:YM}

We would now like to illustrate Eq.~\eqref{eq:ym_theta_sym} by explicitly looking at symmetry properties of the vacua and excitations of Yang-Mills theory. As is well known, $SU(N)$ YM theory on $\mathbb{R}^4$ is asymptotically free and as such becomes strongly coupled at energy scales small compared to the inverse strong scale, $\Lambda^{-1}$. Hence methods of studying the explicit vacuum structure of the theory are limited. Instead we choose to study YM theory on $\mathbb{R}^3\times S^1$, with a circle size of $L$. In this system the vacuum dynamics are calculable via weak coupling methods, specifically in the limit where $NL\Lambda\ll 1$ and center symmetry is preserved. There has been significant evidence \cite{
Unsal:2007vu,Unsal:2007jx,Shifman:2008ja,Shifman:2008cx,Shifman:2009tp,
Cossu:2009sq,Myers:2009df,Simic:2010sv,Unsal:2010qh,Azeyanagi:2010ne,Vairinhos:2011gv,
Thomas:2011ee,Anber:2011gn,Unsal:2012zj,
    Poppitz:2012sw,
    Poppitz:2012nz,Argyres:2012ka,Argyres:2012vv,
    Anber:2013doa,Cossu:2013ora,
    Anber:2014lba,Bergner:2014dua,Bhoonah:2014gpa,Li:2014lza,
    Anber:2015kea,Anber:2015wha,Misumi:2014raa,
    Cherman:2016hcd,Aitken:2017ayq,Anber:2017rch,Anber:2017pak,Anber:2017tug,Anber:2017ezt} that YM depends smoothly on the parameter $NL\Lambda$, and hence it is conjectured that one can recover results for the theory on $\mathbb{R}^4$ in the large $L$ limit. We will begin by briefly reviewing such a system. Those concerned only with our analysis of the vacuum can skip to Sec. \ref{sec:vacua}.

\subsection{Weak-coupling setup} 
\label{sec:YM_setup}   

Consider pure $SU(N)$ Yang-Mills theory on $\mathbb{R}^3\times S^{1}$.   For small $S^1$, it is known that the $\mathbb{Z}_N$ center symmetry is spontaneously broken \cite{Gross:1980br,Weiss:1981ev}, while for large $S^1$ the symmetry is expected to be restored.  The order parameter for the associated phase transition is the expectation value of the trace of powers of 
\begin{align}\label{eq:holonomy}
\Omega(x^\mu)=P\exp\left[i\int_{0}^{L}dx_4\,A_{4}\left(x^\mu,x_4\right)\right]
\end{align}
where we have changed conventions slightly and will henceforth use $\mu,\nu=1,2,3$. At large $L$, $\langle \tr \Omega^k \rangle = 0$ for $k \neq 0 \textrm{ mod } N$, while at small $L$ $\langle \tr \Omega \rangle \neq 0$.  However, if one is not interested in interpreting $S^1$ as a Euclidean thermal circle, this phase transition can be avoided by ``center-stabilizing" deformations.  One example of such a deformation is the addition of $N_F>1$  massive Majorana adjoint fermions with mass $m_{a} \lesssim 1/(NL)$ \cite{Unsal:2010qh}.  Another example is the addition of a double-trace deformation  \cite{Unsal:2008ch}.   With either deformation, it is believed that center symmetry is then preserved for all $L$, with the benefit that at small $L$ the physics becomes analytically calculable. 

We choose to explore the behavior of the symmetries in the center-symmetric phase of the theory that follows from either of deformations referenced above.  At small $L$, where quantum fluctuations become small, the holonomy takes the form   
\begin{align}
\langle \Omega \rangle = \omega^{-(N-1)/2} \mathrm{diag}(1,\omega, \ldots, \omega^{N-1}),\qquad \omega = e^{2\pi i /N}.
\label{eq:center_sym_holonomy}
\end{align}
up to gauge transformations.

We will analyze the theory at distances large compared to $L$, where the system can be described by a 3D effective field theory. From \eqref{eq:holonomy}, the holonomy eigenvalues above imply that (in a standard gauge-fixed sense) $\left\langle A_4 \right\rangle \ne 0$ , which acts as an adjoint Higgs field in the 3D EFT, and breaks the gauge group down to $U(1)^{N-1}$.  The lightest W-bosons have the tree-level mass 
\begin{align}
m_W \equiv \frac{2 \pi}{ NL}.
\end{align}
So when $m_W \gg \Lambda$ --- equivalently, when $NL\Lambda \ll 1$ --- the gauge coupling stops running at the scale $m_W$, and the long-distance 3D effective field theory becomes weakly coupled.  We focus on this tractable limit for the remainder of this paper.

The lightest fields in the 3D effective field theory are the $U(1)^{N-1}$ gauge bosons, the ``photons".  It is useful to note that the associated field strength operators $F_{\mu\nu}^{a}, a =1, \ldots, N-1$, have a gauge-invariant 4D interpolating operator representation given by 
\begin{align}
\label{eq:gi_famu}
F^{a}_{\mu \nu}(x_{\mu})  \sim \frac{1}{N} \frac{1}{L} \int d x_4 \sum_{q=1}^{N-1} \omega^{-q a} \tr \Omega^{q}(x^{\mu}) F_{\mu \nu}(x_4,x_{\mu}) \, ,
\end{align}
with $F_{\mu \nu}$ the 3D part of the $SU(N)$ non-Abelian field strength.  This representation makes clear that the ``color" index can actually be thought of as the discrete Fourier transform of the winding number of a topologically non-trivial state.  

In terms of these fields, the tree-level action of the 3D EFT can be written as 
\begin{align}
\label{eq:euc_action}
S_{\rm tree}  = \frac{L}{4g^2} \int d^3{x} \sum_{a =1}^{N} F^{a}_{\mu \nu} F^{a\mu \nu} \,.
\end{align}
For later notational convenience we have introduced a fictitious $N$th photon in writing this expression.  This extra field can be thought of as the diagonal component of a $U(N)$ field strength, and exactly decouples from the physical adjoint fields in our system.  Using Eq.~\eqref{eq:gi_famu} one can show that center symmetry acts as
\begin{align}
\mathcal{S}: \; F^{a}_{\mu \nu} \to F^{a+1}_{\mu \nu} \,. 
\label{eq:center3d}
\end{align}

In order to analyze the non-perturabtive dynamics of our system, we follow Ref.~\cite{Polyakov:1976fu,Unsal:2008ch} and rewrite \eqref{eq:euc_action} by dualizing the photon, trading $F_{\mu\nu}^{a}$ for a pseudoscalar field $\sigma^{a}$ via the relation 
\begin{align}
F^{a}_{\mu \nu}\equiv\frac{\lambda}{4\pi^2}\epsilon_{\mu\nu\rho}\partial^\rho \sigma^a , \qquad\lambda = g^2 N.
\end{align}
This allows us to rewrite \eqref{eq:euc_action} as
\begin{align}
S_{\rm tree, dual}  = \lambda m_W \int d^3{x}  \sum_{a =1}^{N} (\partial_{\mu}\sigma^a)(\partial^{\mu}\sigma^a) \equiv \lambda m_W \int d^3{x} ~ (\partial_{\mu}\vec{\sigma})^2
\label{eq:perturbative_action}
\end{align}
where we have defined the $N$-component vector of dual photon fields $\vec{\sigma}=(\sigma^1,\ldots,\sigma^N)$.

The dual photons in \eqref{eq:perturbative_action} have no potential to all orders in perturbation theory.  So there is no mass gap in perturbation theory. However, the theory has finite-action field configurations that generate a non-perturbative potential for $\vec{\sigma}$.  In Appendix \ref{app:Disc_syms}, we review the finite-action solutions of this theory with the smallest action.  They come in $N$ distinct types, and are usually called monopole-instantons. They carry topological charge $Q_T = 1/N$, action $S_0=8\pi^2/\lambda$, and carry magnetic charges associated to the simple (co-)roots $\vec{\alpha}_a$ of the affine extension of the $\mathfrak{su}(N)$ Lie algebra. For more details on the non-perturbation solutions and their transformations under the symmetries of the theory, see Appendix \ref{app:Disc_syms}. 

As explained in \cite{Unsal:2008ch}, summing over the contributions of the monopole-instanton solutions to the path integral using a dilute-gas approximation (which is well-justified when $NL\Lambda  \ll 1$) produces a potential for the dual photons, so that 
\begin{align}
S_{\rm \vec{\sigma}} = \int d^3{x}\, \left[ \lambda m_W (\partial_{\mu} \vec{\sigma})^2  + V(\vec{\sigma}) \right]
\label{eq:YM_EFT}
\end{align} 
where the non-perturbative potential is given by
\begin{align}
V\left(\vec{\sigma}\right) &= -\frac{A}{\lambda^{2}}m_{W}^{3}e^{-S_0}\sum_{a=1}^{N}\cos\left[\vec{\alpha}_{a}\cdot\vec{\sigma}+\frac{\theta}{N}\right]+ \ldots.
\label{eq:YM_potential}
\end{align}
The ``$\ldots$'' represent higher order contributions which we will neglect here. Here, $A>0$ is an $\mathcal{O}(1)$ scheme-dependent dimensionless constant which will not be important in what follows. The monopole-generated  potential depends on the $\theta$ angle because the monopole-instantons have non-vanishing topological charge.

We now show how the YM symmetry group in \eqref{eq:ym_theta_sym} acts in the effective field theory associated to \eqref{eq:YM_EFT}

\subsection{Extrema and symmetries as a function of $\theta$}
\label{sec:vacua}
We now begin our analysis of the vacuum structure of \eqref{eq:YM_EFT}, with the leading order-potential explicitly shown in \eqref{eq:YM_potential}.  The dual photon fields live in the weight lattice of $\mathfrak{su}(N)$.  The potential has $N$ extrema in the unit cell of the weight lattice at
\begin{equation}
\vec{\sigma}_{k}=\frac{2\pi k}{N}\vec{\rho},  \qquad \mathrm{with}\qquad\vec{\rho}\equiv\sum_{i=1}^{N-1}  \vec{\mu}_{i}
\label{eq:YMextrema}
\end{equation}
where $k=0,\ldots,N-1$. Here $\vec{\mu}_{i}$ are the $SU(N)$ fundamental root vectors, and satisfy $\vec{\alpha}_{i}\cdot\vec{\mu}_{j}=\delta_{ij}$, and $\vec{\rho}$ is the Weyl vector satisfying  $\vec{\alpha}_{i}\cdot\vec{\rho}=1$ for $i = 1, \ldots, N-1$ and $\vec{\alpha}_{N}\cdot\vec{\rho}= 1 - N$.   For example, in a basis where $(\alpha_{a})_b = \delta_{a,b} - \delta_{a+1,b}, \, 1\le a < N$, $\vec{\sigma}_{k}$ takes the form
\begin{align}
\vec{\sigma}_{k} = \frac{2\pi k}{N} (N, N-1, \ldots, 2,1) \, .
\label{eq:weyl_vector}
\end{align}
The non-perturbative 3D  energy density evaluated at each of these extrema is
\begin{equation}
V_{k} \equiv V\left(\vec{\sigma} = \vec{\sigma}_k \right)=-N\frac{A}{\lambda^{2}}m_{W}^{3}e^{-S_0}\cos\left(\frac{2\pi k+\theta}{N}\right) + \mathcal{O}(e^{-2S_0}) \,.
\label{eq:pot_vals_tnm}
\end{equation}

For any given $\theta$, the integer $k$ labeling the globally-stable ground state is determined by minimizing \eqref{eq:pot_vals_tnm}.   The metastable states of the system will correspond to the subset of  extrema with positive curvature in all directions in $\vec{\sigma}$ space.  
On any fixed branch, the physics is periodic in $2\pi N$.  However, the $k$ that minimizes $V_{k}$ depends on $\theta$.  Thus, just from the form of \eqref{eq:pot_vals_tnm}, one can see that as $\theta$ varies in $[0,2\pi)$, the value of $k$ associated with the minimal energy extremum will change in such a way that the physics of the complete system in its ground state has a $\theta$ periodicity of $2\pi$.  However, the observables are non-analytic functions of $\theta$, which is associated with jumps in the value of $k$ which minimize the ground state energy density.  This is consistent with Witten's conjectured picture \cite{Witten:1978bc,Witten:1998uka} for the $\theta$-dependence of YM theory.  Earlier discussions of how $2\pi$ periodicity emerges in the present context were presented in e.g. Refs.~\cite{Unsal:2008ch,Thomas:2011ee,Unsal:2012zj,Poppitz:2012sw,Poppitz:2012nz,Bhoonah:2014gpa,Anber:2017rch}. 

Let us now understand how center and coordinate reflection symmetries act on the extrema of \eqref{eq:YMextrema}.  To do this, it is useful to work out how these transformations act in compactified YM theory more generally, see Appendix \ref{app:Disc_syms}, and also Ref.~\cite{Aitken:2017ayq}.  Here we will focus on reflections of the compactified coordinate $\mathcal{R}$, charge conjugation $\mathcal{C}$, and (0-form) center transformations $\mathcal{S}$.  The effective field theory on $\mathbb{R}^3 \times S^1$ is built from the dual photon fields $\sigma_a$, and the action of these transformations 
which follows from \eqref{eq:gi_famu} and \eqref{eq:center3d}, 
 is
\begin{align}
\mathcal{S}: \sigma_a &\to \sigma_{a+1}\\
\mathcal{C}: \sigma_a &\to - \sigma_{N-a+1}\\
\mathcal{R}: \sigma_a &\to \begin{cases}
						  \sigma_{N-a+1}\,, & \theta = 0 \\
						  \sigma_{N-a+1} -\frac{2\pi(N-a+1)}{N}\,, & \theta = \pi 
					\end{cases}
\end{align}
Looking at the form of the effective action \eqref{eq:YM_EFT}, it is clear that $\S$ and $\C$ are symmetries for any $\theta$, as one would expect.  The $\mathcal{R}$ coordinate-reflection transformation is a symmetry only if $\theta = 0$ or $\theta = \pi$.  Note that when acting on $\vec{\alpha}_{a} \cdot \vec{\sigma}$ at $\theta=\pi$, the reflection symmetry transformation gives 
\begin{align}
\mathcal{R} : (\vec{\alpha}_{a} \cdot \vec{\sigma}) \to -\vec{\alpha}_{N-a} \cdot \vec{\sigma} -\frac{2\pi}{N} \,.
\end{align}
The resulting shift in the phase of monopole operators is necessary because a coordinate reflection must be accompanied by a $2\pi$ shift in the $\theta$ angle to be a symmetry of the theory.

One can now easily work out the symmetry group. To do so, consider the action of the symmetry transformations on an operator of the form $e^{i\sigma_a}$.   It can be checked that $\C^{-1} \S \C = \S^{-1}$, corresponding to a $D_{2N}$ symmetry group, just as one would expect from the general arguments in Sec.~\ref{sec:symmetry}.  For the $\R$ and $\S$ symmetries, we obtain
\begin{align}\label{eq:dYM_algebra}
\R^{-1} \S \R = \begin{cases}
			\S^{-1} \,, &\theta = 0 \\
			\omega \S^{-1} \,, &\theta = \pi \,.
			\end{cases}
\end{align}
This corresponds to a $D_{2N}$ group for $\theta = 0$, and a $D_{4N}$ group for $\theta = \pi$.  As in in our discussion of the $T_{N}$ model, for even $N$ we interpret the $\theta = \pi$ commutation relations in \eqref{eq:dYM_algebra} to imply the existence of a mixed 't Hooft anomaly between center and time-reversal symmetries, while for odd $N$ we interpret them to imply a global inconsistency between these symmetries.  
In total, we find precisely the expected 0-form symmetries of \eqref{eq:ym_theta_sym}, reproduced here for convenience
\begin{align}
G^{\rm discrete}_{\rm YM} = \begin{cases}
D_{2N} \times \mathbb{Z}_2 \times \mathbb{Z}_2
& \theta=0\; \mathrm{mod}\; 2\pi \\
D_{4N} \times \mathbb{Z}_2 \times \mathbb{Z}_2
& \theta=\pi\; \mathrm{mod}\; 2\pi \\
D_{2N} & \text{otherwise}.
\end{cases}
\label{eq:ym_theta_sym_2}
\end{align}
Note that a benefit of our approach we get a simple picture for how the mixed center-$CP$ 't Hooft anomaly of Ref.~\cite{Gaiotto:2017yup} arises (as a central extension of the symmetry group, just like in toy QM examples).  Moreover, given that we work in a regime where the dynamics is calculable, we can fully determine the vacuum structure. On the other hand, the general nature of the considerations of Ref.~\cite{Gaiotto:2017yup} have their own benefits.  In particular, they are valid regardless of the strength of the coupling in the system.  We explore further features of the vacuum structure of \eqref{eq:YM_potential} and higher order corrections in a companion paper \cite{Aitken:2018mbb}.
 
Turning back to the symmetry transformations of the extrema of the potential,  we find that $\R$ acts as
\begin{align}
\mathcal{R} :    \vec{\sigma}_{k} \to 
\begin{cases}  
\vec{\sigma}_{-k} & \theta = 0  \\
  \vec{\sigma}_{-k+1} & \theta = \pi.
 \end{cases}
\end{align}
while the center transformation rule is 
\begin{equation}
\S\::\: \vec{\sigma}_{k} \to \vec{\sigma}_{k} + \frac{2\pi k }{N}\vec c
\label{eq:center_trans_dym}
\end{equation}
where the $N$-vector $\vec{c}$ obeys the relations 
\begin{equation}
\vec{\alpha}_{a} \cdot \vec c =\begin{cases}
-N & a = 1\\
0 & 1 < a <N\\
N & a=N
\end{cases}
\end{equation}
 For example, in the basis of Eq.~\eqref{eq:weyl_vector}, $\vec c = \left(1,1,\ldots,1,1-N\right)$.   The condition that $\S^N\cdot \vec{\sigma}_k = \vec{\sigma}_k$ is related to the periodicity of the $\sigma_{a}$ fields and the quantization of the coefficient of $\vec{c}$ in \eqref{eq:center_trans_dym}.   

\section{Summary}
\label{sec:outlook}
We have examined the global symmetries and ground state properties of $SU(N)$ YM theory as a function of the topological $\theta$ angle.  The global symmetries were argued to include non-Abelian discrete groups --- specifically, dihedral groups --- for all $\theta$ when $N\ge3$ due to a non-commutativity between center symmetry and charge conjugation.  We then examined the vacuum structure of YM theory as a function of $\theta$.  First, we warmed up by considering a simple quantum mechanics example whose symmetries also include dihedral groups. We then used the technique of adiabatic circle compactification of YM theory on $\mathbb{R}^3\times S^1$ to illustrate the symmetry structure and some ground state properties in a systematically calculable setting. 


\section*{Acknowledgments}

We are grateful to  S. Gukov, K.~Jensen, M.~Shifman, T.~Sulejmanpasic, and  L.~Yaffe for helpful conversations. We are especially indebted to L. Yaffe for his extensive comments and advice on a draft of the paper. K.~A. is supported by the U.S.~Department of Energy under Grant No.~DE-SC0011637. A.~C. and M.~\"U.  thank the KITP for its warm hospitality as part of the program ``Resurgent Asymptotics in Physics and Mathematics'' during the final stages of the research in this paper. Research at KITP is supported by the National Science Foundation under Grant No. NSF PHY11-25915.  A.~C. is also supported by the U. S. Department of Energy via grants DE-FG02-00ER-41132, while M.~\"U. is supported U. S. Department of Energy grant DE-FG02-03ER41260. 


\appendix

\section{Path integral formulation of the $T_N$ model}
\label{appendix:path_integral_QM}

In this appendix we consider the path integral description of anomalies and global inconsistency conditions in the quantum-mechanical $T_N$ model.  Our exposition is based on \cite{Kikuchi:2017pcp}, see also \cite{Gaiotto:2017yup}. 

A mixed 't Hooft anomaly is when $G=G_1\times G_2$ and gauging of one of the symmetries results in the loss of the other. This motivates turning on a background gauge field associated to $\S$.  $\S$ is a discrete 0-form 
$\mathbb Z_N$ symmetry and gauging it involves coupling the $T_N$ model to a topological field theory \cite{Banks:2010zn,Kapustin:2014gua}. For this, it turns out that it is most efficient to work with two background gauge fields $A,B$ where $A$ is 1-form and $B$ is 0-form,   related by a constraint induced by some Lagrange multiplier field $F$. The action of the $T_N$ model with background fields associated to $\S$ is
\begin{align}
S_{T_N}(A,B;g,\theta,\tftp) = \frac{1}{g^2} \int &\left[\frac{1}{2}(dq+A)^2-\cos\left(Nq+B\right)\right] 
-\frac{i \theta}{2\pi} \int (dq +A) \\
&+ i \int F \wedge(d B - N A) + i \tftp \int A
\label{eq:TNgauged}
\end{align}  
and the partition function is
\begin{align}
Z_{T_N}(A,B; g, \theta, \tftp) = \int d[q] d[F] e^{-S_{T_N}(A,B;g,\theta,\tftp)}.
\end{align}
Note that integrating out $F$ enforces the on-shell identity $d B = N A$.  The (background) 0-form gauge transformation properties are
\begin{align}
q &\to q - \lambda \\
B &\to B + N \lambda\\
A &\to A + d \lambda\\
F &\to F.
\end{align}
One can check that the action is invariant under these gauge transformations so long as the coefficient of the 1D Chern-Simons term $\tftp$ is an integer.   The integer $\tftp$ can be interpreted as a hidden parameter in the theory in addition to the obvious parameters $g, \theta$, and to define the theory for any value of the background fields we must specify all \emph{three} parameters $g, \theta, \tftp$.

The fact that the action \eqref{eq:TNgauged} is gauge invariant means that there is no direct 't Hooft anomaly for $\S$.  However, since the system has additional discrete symmetries at $\theta=0$ and $\pi$, these points of parameter space are potentially problematic and should be checked for mixed 't Hooft anomalies. 

At $\theta = 0$, it is easy to check that as long as $2\tftp = 0 \textrm{ mod } N$, $\C$ and $\T$ are symmetries with the transformation rules 
 \begin{align} 
 \C: \; &\{t\to +t, q \to -q, A \to -A, B \to -B, F\to -F\} \\
 \T:\;  &\{t\to -t, q \to +q, A \to -A, B \to +B, F\to -F\} \, .
 \end{align}
At $\theta = \pi$, on the other hand, the $\C$ and $\T$ transformations are symmetries so long as $2\tftp-1 =  0 \textrm{ mod } N$.

The $\theta =0$ symmetry condition $2 \tftp = 0 \textrm{\, mod\, } N$ can always be satisfied without violating the integrality of $\tftp$ by setting $\tftp = 0$ (or $N/2$ for $N$-even). But the $\theta = \pi$ symmetry condition $2\tftp-1 = 0  \textrm{\, mod\, } N$ has much stronger consequences.  For even $N$, it cannot be satisfied at all with integer $\tftp$.  This can be interpreted  as a mixed 't Hooft anomaly between the discrete shift symmetry and both $\C$ and $\T$. Consequently, either one of the $\C, \T$ symmetries must be spontaneously broken, or the shift symmetry $\S$ must be broken.  

For odd $N$, the $\theta = \pi$ symmetry condition can be satisfied by e.g. $\tftp = (N-1)/2$, so one can preserve $\C$ and $\T$.  This means that with appropriate choices of $\tftp$ one can preserve $\C$ and $\T$ at either $\theta = 0 $ and $\theta = \pi$.  But the values of the discrete parameter $\tftp$ necessary to keep $\C$ and $\T$ symmetries at $\theta = 0$ and $\theta = \pi$ are not the same.  So if one \emph{defines} the theory with a fixed choice of $\tftp$ which preserves $\C, \T$ symmetries at $\theta = 0$, then one cannot trivially maintain all three discrete symmetries $\C, \T, \S$ at $\theta = \pi$.  The simplest possibility is that one of these symmetries should be spontaneously broken at $\theta = \pi$. In this sense there is always a global inconsistency between the $\C, \T$ symmetries and the $\S$ symmetry for any $N > 1$, but there is the slightly stronger condition of a mixed 't Hooft anomaly for even $N$.

Of course, this is a simple quantum mechanical system, so one can back up the claims of the preceding paragraphs and verify the degeneracy of the ground states by either diagonalizing the Hamiltonian numerically or solving it semi-classically.  Indeed, at $\theta =\pi$ time-reversal/charge conjugation breaks spontaneously for all $N>1$.

\section{Representations of the dihedral group }
\label{app:Reps_dihedral}

In order to find the decomposition of states in terms of irreducible representation, we calculate the character associated with the conjugacy classes of $D_{2N}$. Recall that the character of a group element $g$ in a representation $R$ is given by $\chi_{R}\left(g\right)=\text{tr}D_R(g)$, with $D_R(g)$ the group element $g$ in representation $R$. Expressing this character in terms of characters of the irreducible representations via orthogonality relations then allows us to find the decomposition of $R$. 

\subsubsection*{Even $N$: $T_{N=2k}$ Model}

To find the characters we want to find the general form of the $N$-dimensional representation, $R$, corresponding to how the $N$ translation eigenstates (Bloch states) $\ket{k}$ transform under $\S=s$ and $\mathcal{C}=r$. For example, in the $N=4$ case , which corresponds to $D_{8}$ and $k=2$, using \eqref{eq:tn_S_trans} and \eqref{eq:tn_C_trans} give
\begin{equation}
r=\left(\begin{array}{cccc}
\omega^{-1}\\
 & \omega^{-2}\\
 &  & \omega^{-3}\\
 &  &  & \omega^{-4}
\end{array}\right),\qquad s=\left(\begin{array}{cccc}
 &  & 1\\
 & 1\\
1\\
 &  &  & 1
\end{array}\right).
\end{equation}
Generalizing the form of $r$ and $s$ above, it is straightforward to find the characters for arbitrary $N$. Note that nonzero contributions to a transformation's character correspond to states which are mapped back to themselves under such a transformation. Identifying such states will often be a useful tool in finding characters for arbitrary $N$. Obviously the identity element has character $N$. The generalization of $r$ to arbitrary $N$ is a diagonal matrix with all $N$th roots of unity, and as such the $\text{tr}r=0$. This also holds for $r^{m}$ for any $m=1,\ldots,N-1$, since $r^{m}$ correspond to the $N/\text{gcd}\left(N,m\right)$th roots of unity.

We see that $s$ maps precisely two minima back to themselves, and so it will have character $2$. This holds for arbitrary $N=2k$ since there will always be two elements where $N-p\,\text{mod}N=p$, namely $\frac{N}{2}=k$ and $N$. For the $N=4$ case, $sr^{2}$ also maps two minima back to themselves and hence also has character $2$. This follows more easily from the fact all members of a conjugacy class have the same character, and hence if the character of $s$ is $2$, so too must $\left\{ sr^{2b}\right\} $. However, $sr$ and $sr^{3}$ have character $0$, since the two nonzero diagonal elements of $s$ will always pick out elements of $r^{2m+1}$ which are $\pi$ out of phase on the unit circle (i.e. $\omega^{-2}$ and $\omega^{-4}$ for $N=4$ and $m=0$ case). Explicitly, the nonzero elements correspond to the $N$th and $\frac{N}{2}$th position, and the $N$th position is always $1^{2m+1}=1$ and the $\frac{N}{2}$th position is $\omega^{-\left(N/2\right)\left(2m+1\right)}=\omega^{-Nm-N/2}=\omega^{-N/2}=-1$. Hence the characters of the conjugacy classes, (\ref{eq:d_even cc size tnm}),
for arbitrary $N=2k$ are\\
\begin{equation}
\chi_{\text{even}}^{\theta=0}=\{N,\underset{k-1}{\underbrace{0,0,\ldots,0}},0,2,0\}.
\end{equation}

The general character table for $D_{2N}$ is given in e.g.  \cite{Ramond:2010zz}. We can use character orthogonality to find the decomposition of a representation. Namely, for a given representation $R$ with characters $\chi_{R}$, the number of a given irrep. $R_{1}$ with characters $\chi_{R_{1}}$ is given by
\begin{equation}
\frac{1}{n}\sum_{i=1}^{K}s_{i}\chi_{R}\chi_{R_{1}}=\text{\# of a \ensuremath{R_{1}} irrep. in arbitrary representation}\label{eq:char ortho tnm}
\end{equation}
where $n$ is the number of elements in the group (i.e. $2N$ for $D_{2N}$), $K$ is the number of conjugacy classes, and $s_{i}$ is the size of the $i$th conjugacy class. Since the only nonzero terms in the character table are those corresponding to classes $\left\{ 1\right\} $ and $\left\{ sr^{2b}\right\} $, it is straightforward to perform the projections and find the decomposition of \eqref{eq:d_even th=0}.

At $\theta=\pi$, the $N$ low lying states transform under the action of $D_{4N}$ group elements (so now $k=N$). The conjugacy classes and number of elements in each class are again given by \eqref{eq:d_even cc tnm} and \eqref{eq:d_even cc size tnm}, but now $2N=M=k$. The representation here is slightly more complicated because we need to find the $N$-dimensional representation of $D_{4N}$. However, a natural definition is motivated by our definition $\tilde{\S}=\omega^{1/2}\S$, so we can take $\tilde{r}=\omega^{1/2}r$ where $r$ is the $N$-dimensional representation of $D_{2N}$. The form of $s$ follows from \eqref{eq:tn_C_trans}. For example, for $N=4$ the $N$-dimensional representation given by
\begin{equation}
\tilde{r}=\left(\begin{array}{cccc}
\omega^{-1/2}\\
 & \omega^{-3/2}\\
 &  & \omega^{-5/2}\\
 &  &  & \omega^{-7/2}
\end{array}\right),\qquad\tilde{s}=\left(\begin{array}{cccc}
 &  &  & 1\\
 &  & 1\\
 & 1\\
1
\end{array}\right)
\end{equation}
does the trick.

Once more, generalization to arbitrary $N$ is not difficult. The identity again has $\chi_{I}=N$, meanwhile all $\tilde{r}$ still have $\chi=0$ (since shifting the roots of unity uniformly by $\omega^{1/2}$ does not change their cancellation) with the exception of $\tilde{r}^{2N}=-1$, which has character $-N$. Now, $s$ maps $\ket p\to\ket{N-p+1}$, and hence no elements are mapped back to themselves corresponding to zero character. Multiplication of $\tilde{s}$ by any combination of $\tilde{r}$ does not change the location of nonzero elements, so any combination $\tilde{s}\tilde{r}^{i}$ for $i=1,\ldots,2N$ also has zero trace. Thus the characters are
\begin{equation}
\chi_{\text{even}}^{\theta=\pi}=\{N,\underset{N-1}{\underbrace{0,0,\ldots,0}},-N,0,0\}.
\end{equation}

Once more, for the decomposition it is only the nonzero components we should worry about, this time corresponding to the 1-element conjugacy classes $\left\{ 1\right\} $ and $\left\{ r^{2N}\right\} $. Using Table \ref{tab:D2N_char_tab}, $c_{2N\left(2m-1\right)} =-1$, and $c_{2N\left(2m\right)}=1$ the decomposition in terms of irreducible characters yields \eqref{eq:d_even th=pi}.

\subsubsection*{Odd $N$: $T_{N=2k+1}$ Model}

The $N$-dimensional representation follows in a very similar manner as before. For example, $N=5$ yields
\begin{equation}
r=\left(\begin{array}{ccccc}
\omega^{-1}\\
 & \omega^{-2}\\
 &  & \omega^{-3}\\
 &  &  & \omega^{-4}\\
 &  &  &  & \omega^{-5}
\end{array}\right),\qquad s=\left(\begin{array}{ccccc}
 &  &  & 1\\
 &  & 1\\
 & 1\\
1\\
 &  &  &  & 1
\end{array}\right).
\end{equation}

The characters for $r^{m}$ with $m=1,\ldots,N-1$ follow similarly. The primary difference here is the fact $s$ will only bring a single element back to itself, and this is unchanged when multiplying by any power of $r$ since the $N$th diagonal position will always be $\omega^{-Nm}=1$. Hence the characters are given by:
\begin{equation}
\chi_{\text{odd}}^{\theta=0}=\{N,\underset{k}{\underbrace{0,0,\ldots,0}},1\}.
\end{equation}
The characters for arbitrary odd $N$ are given in Table \ref{tab:D2N+1 char tab tnm}. Using the orthogonality of characters, (\ref{eq:char ortho tnm}), for the nonzero elements we find \eqref{eq:d_odd th=0}.

For $\theta=\pi$ we found a global inconsistency condition which implied the group was centrally extended to $D_{4N=2(4k+2)}$. Note this has switched us from conjugacy classes of the form (\ref{eq:d_even cc tnm}) with $M=2N$ instead of those of (\ref{eq:d_odd cc tnm}), so we should use the character table of Table \ref{tab:D2N_char_tab}. Building an $N$ dimensional representation for $D_{4N}$ from the $N$-dimensional representation of $D_{2N}$ follows in an analogous manner as before. For $N=5$,
\begin{equation}
\tilde{r}=\omega^{1/2}\left(\begin{array}{ccccc}
\omega^{-1}\\
 & \omega^{-2}\\
 &  & \omega^{-3}\\
 &  &  & \omega^{-4}\\
 &  &  &  & \omega^{-5}
\end{array}\right),\qquad s=\left(\begin{array}{ccccc}
 &  &  &  & 1\\
 &  &  & 1\\
 &  & 1\\
 & 1\\
1
\end{array}\right).
\end{equation}

Again, the identity and $\tilde{r}^{2N}$ yield $N$ and $-N$, respectively. We see from the above representation, $s$ will map one element back to itself. This generalizes for $sr^{2b}$ with $b=1,\ldots,2k+1$ since this element will always be that which corresponds to $\omega^{1/2-\left(N+1\right)/2}$ and
\begin{align*}
\omega^{\left[1/2-\left(N+1\right)/2\right]m} & =\omega^{-Nm/2}=\begin{cases}
1 & m\text{ even}\\
-1 & m\text{ odd}
\end{cases}.
\end{align*}
Hence, the characters of the conjugacy classes are
\begin{equation}
\chi_{\text{odd}}^{\theta=\pi}=\{N,\underset{N-1}{\underbrace{0,0,\ldots,0}},-N,1,-1\}.
\end{equation}
The character orthogonality takes slightly more work, but follows in a similar manner. Using Table \ref{tab:D2N_char_tab}, $\left(-1\right)^{N}=\left(-1\right)^{2k+1}=-1$, $c_{2N\left(2m-1\right)}=-1$, and $c_{2N\left(2m\right)} =1$, the decomposition of \eqref{eq:d_odd th=pi} is found.

\section{Discrete symmetries of YM on $\mathbb{R}^3 \times S^1$}\label{app:Disc_syms}
In this appendix, we investigate the discrete symmetries of deformed YM in greater detail and justify why $CP$ is indeed the symmetry which interchanges extrema with the same $V_k$. Since the potential from which we derive these symmetries in a result of a non-perturbative dilute gas summation of monopole-instanton solutions, this necessarily requires a closer investigation of how such solutions transform under discrete symmetries. For completeness, we first review the monopole-instanton solutions of deformed YM. We then investigate how the degenerate extrema are related to one another and see how discrete symmetries act on these solutions.

\subsection*{Monopole-instanton solutions}
Monopole-instanton solutions are found using the usual BPS trick on the Euclidean action, \eqref{eq:euc_action} \cite{Anber:2011gn}. For simplicity, suppose $N=2$.  Then we can express the action in terms of the chromo-electric and chromo-magnetic fields from the non-Abelian field strengths (recall, $x^{4}$ is the compact
direction and $\mu,\nu=1,2,3$)
\begin{equation}
E_{\mu}^{a}=F_{\mu 4}^{a}=D_{\mu}^{ba}A_{4}^{b}\qquad B_{\mu}^{a}=\frac{1}{2}\epsilon_{\mu\nu\rho}F_{\nu\rho}^{a}\label{eq:e and b fields tnm}
\end{equation}
with $D_{\mu}^{ba}=\partial^{\mu}\delta^{ab}+\epsilon^{abc}A_{\mu}^{c}$ and $F_{\mu\nu}^{a}=\partial_{i\mu}A_{\nu}^{a}-\partial_{\nu}A_{\mu}^{a}+\epsilon^{abc}A_{\mu}^{b}A_{\nu}^{c}$
and so \eqref{eq:euc_action} with a nonzero $\theta$ term  becomes
\begin{align}
S_{\text{tree}} = \frac{L}{2g^{2}}\int d^{3}x\,\left(E_{\mu}^{a}\mp B_{\mu}^{a}\right)^{2}+\left(\frac{i\theta L}{16\pi^{2}}\pm\frac{L}{g^{2}}\right)\int d^{3}x\,E_{\mu}^{a}B^{a\mu}\label{eq: e and b action tnm}
\end{align}
where the top (bottom) corresponds to the monopole (anti-monopole) solution. We see the monopole and anti-monopole then satisfy 
\begin{equation}
E_{\mu}^{a}=\pm B_{\mu}^{a}\qquad\Leftrightarrow\qquad F_{MN}^{a}=\pm\tilde{F}_{MN}^{a}.\label{eq:duality condition tnm}
\end{equation}
with $M,N=1,2,3,4$. The monopole solutions carry magnetic and topological charge, defined
by
\begin{equation}
Q_{T}\sim\int d^{3}x\,E_{\mu}^{a}B^{a\mu},\qquad Q_{M}^{a}\sim\int d^{2}x\,\hat{n}^{\mu}B_{\mu}^{a}.\label{eq:qm qt defs tnm}
\end{equation}

The standard $\mathbb{\mathbb{R}}^{4}$ monopole solutions which arise from (\ref{eq:duality condition tnm}) can be constructed such that they are independent of one spacetime coordinate, and thus have the properties of pseudo-particles (co-dimension one). When we dimensionally reduce from $\mathbb{\mathbb{R}}^{3}\times S^{1}$ to $\mathbb{\mathbb{R}}^{3}$, so long as we choose the compactified direction to correspond to the direction in which our monopole solutions are independent, we will end up with a ``monopole-instanton solution'' (co-dimension zero). Monopole/anti-monopole (instanton) solutions derived in this way are $x_{4}$ independent. It is also possible to find $x_4$ dependent solutions with the same action by allowing $\pm 1$ units of KK momentum \cite{Lee:1997vp,Kraan:1998pm}. This results in a total of $N$ monopole solutions with action $S_{0}\equiv8\pi^{2}/g^{2}N$, magnetic charge $\vec{\alpha}_{a}$, and topological charge $1/N$, and $N$ anti-monopoles with opposite magnetic and topological charges. 

\subsection*{Monopole transformation properties}

\begin{table}[htb!]
\begin{centering}
\begin{tabular}{ccccc}
\hline 
Transformation & Definition & $Q_{M}$ & $Q_{T}$ & Holonomy Eigenvalues\tabularnewline
\hline 
$\P_{x_{\mu}}$ & $x_{\mu}\to-x_{\mu}$, $A_{\mu}\to-A_{\mu}$ & $-$ & $-$ & unchanged\tabularnewline
$\P_{t}$ & $x_{3}\to-x_{3}$, $A_{3}\to-A_{3}$ & $-$ & $-$ & unchanged\tabularnewline
$\R$ & $x_{4}\to-x_{4},A_{4}\to-A_{4}$ & $+$ & $-$ & $a\to N-a+1$\tabularnewline
$\C$ & $A_{M}\to-A_{M}$ & $-$ & $+$ & $a\to N-a+1$\tabularnewline
$\C\P\R$ & $x^{M}\to-x^{M}$ & $+$ & $+$ & unchanged\tabularnewline
\hline 
\end{tabular}
\par\end{centering}
\caption{Various discrete symmetries and how they transform $Q_{M}^{a}\sim\int d^{2}x\,\hat{n}^{\mu}B_{\mu}^{a}$ and $Q_{T}\sim\int d^{3}x\,E_{\mu}^{a}B_{\mu}^{a}$. A ``$+$'' sign denotes the charge is unchanged under the corresponding transformation, while ``$-$'' sign indicates a flip in sign.\label{tab:Various-discrete-symmetries tnm}}
\end{table}

We now consider how the monopole and anti-monopole solutions are changed under discrete transformations. This will allow us to understand how the Abelian $\sigma_a$ fields transform and ultimately the behavior of $\vec{\sigma}_k$ under these symmetries. 

The monopole and anti-monopole solutions are flipped under a parity transformation in $\mathbb{\mathbb{R}}^{3}$, which we will denote $\P_{x_{\mu}}$. This takes $x_{\mu}\to-x_{\mu}$ and $A_{\mu}\to-A_{\mu}$ which flips the $E$-field but not the $B$-field. However, because of the $\hat{n}^{\mu}$ in the definition of $Q_{M}^{a}$, (\ref{eq:qm qt defs tnm}), which must also flip under $\P_{x_{\mu}}$, this transformation\emph{ does }flip the magnetic charge. Note that a flip of the magnetic charge of the monopoles is equivalent to a transformation of $\vec{\sigma}\to -\vec{\sigma}$. Hence, since both magnetic and topological charge are flipped, this amount to an interchange of monopoles and anti-monopoles. Since the $\theta$-term is proportional to $Q_{T}$, this is a symmetry at only $\theta=0$ and $\theta=\pi$. At $\theta=0$, the invariance is trivial since the topological charge has no effect on the path integral. However, at $\theta=\pi$, the symmetry must be accompanied by a $2\pi$ shift of $\theta$. We can implement such a shift via our $\sigma_a$ variables by defining the action of $\P_{x_{\mu}}$ to be $\theta$-dependent
\begin{align}
\P_{x_{\mu}} :    \sigma_a \to 
\begin{cases}  
-\sigma_a & \theta = 0  \\
  -\sigma_a+\frac{2\pi (N-a+1)}{N} & \theta = \pi.
 \end{cases}
\end{align}

Similarly, consider the parity transformation in a single direction
\footnote{It is tempting to identify this direction as ``time'' to match onto the existing literature. But the considerations here can be phrased in Euclidean space, and all one needs to derive e.g. anomalies is to consider reflections which involve an odd number of directions.  So an identification of $x_4$ with ``time" is possible but not necessary. In particular we find it helpful to think of the $x_4$ direction as a spatial one.  } 
of $\mathbb{\mathbb{R}}^{3}$, which we will take to be $x_{3}$ and denote $\P_{t}$. This takes $x_{3}\to-x_{3}$ and $A_{3}\to-A_{3}$ and hence flips $B_{1}$, $B_{2}$, and $E_{3}$, but leaves the other components of the electric and magnetic field untouched. Since this flips each of the three terms showing up in the topological charge, the net effect is to flip the total topological charge. Additionally, since this flips $\hat{n}_{3}\to-\hat{n}_{3}$, this also flips all three terms showing up in the magnetic charge, and hence takes $Q_{M}^{a}\to-Q_{M}^{a}$.  Thus the net effect of $\P_{t}$ is identical to that of $\P_{x_{\mu}}$, just as one would expect.  We will collectively refer to the two non-compact parity transformations as $\P$. 

Charge conjugation takes $A_{M}\to-A_{M}$, which from (\ref{eq: e and b action tnm}) flips both the electric and magnetic fields.  As such, the magnetic charges of the monopoles are flipped but the topological charges are unchanged. The symmetry thus leaves the $\theta$-term untouched, and hence this symmetry persists for all $\theta$. 
However, from our definition of the holonomy in \eqref{eq:center_sym_holonomy}, $\Omega$ is also affected charge conjugation. More specifically, charge conjugation has the net effect of rearranging the holonomy eigenvalues. In order to leave the theory unchanged, we define charge conjugation to be accompanied by rearrangement of the holonomy eigenvalues so that the net effect of the transformation is to leave the holonomy unchanged (see \citep{Aitken:2017ayq} for more details). At the level of the monopoles, the rearranging of said eigenvalues interchanges monopole labels as $a\to N-a+1$. The combined effect of rearranging labels and flipping the charge means charge conjugation acts on $\sigma_a$ as $\sigma_a\to -\sigma_{N-a+1}$.

Finally, consider the transformation which takes $x_{4}\to L-x_{4}$ and $A_{4}\to-A_{4}$, which we call $\R$. From (\ref{eq:e and b fields tnm}), this flips the $E$-field but not the $B$-field, and hence takes $Q_{M}^{a}\to Q_{M}^{a}$ and $Q_{T}\to-Q_{T}$. However, since a flip in the compact direction transforms $A_{4}$, it will also affect the holonomy in the same way that charge conjugation acted. Hence, we will also define the $\R$ transformation to come with $a\to N-a+1$ relabeling \citep{Aitken:2017ayq}. As with the $\P$ transformations, since $\R$ flips the topological charge we must accompany the transformation at $\theta=\pi$ with an appropriate shift,
\begin{align}
\R :    \sigma_a \to 
\begin{cases}  
\sigma_{N-a+1} & \theta = 0  \\
  \sigma_{N-a+1}-\frac{2\pi (N-a+1)}{N} & \theta = \pi.
 \end{cases}
\end{align}

\newpage

\bibliographystyle{JHEP}
\bibliography{small_circle}

\end{document}